# Smart HAADF-STEM scanning strategy for local measurement of strain at the nanoscale

V. Prabhakara[1,2,3], D. Jannis[1,3], G. Guzzinati[1,3], A. Béché[1,3], H. Bender[2] and J. Verbeeck[1,3]

1. EMAT, University of Antwerp, Groenenborgerlaan 171, 2020, Antwerp, Belgium.

2. Imec, Kapeldreef 75, 3001, Leuven, Belgium.

3. NANOlab Center of Excellence, University of Antwerp, Belgium

**Abstract:**

Lattice strain measurement of nanoscale semiconductor devices is crucial for the semiconductor industry as strain substantially improves the electrical performance of transistors. High resolution scanning transmission electron microscopy (HR-STEM) imaging is an excellent tool that provides spatial resolution at the atomic scale and strain information by applying Geometric Phase Analysis or image fitting procedures. However, HR-STEM images regularly suffer from scanning distortions and sample drift during image acquisition. In this paper, we propose a new scanning strategy that drastically reduces artefacts due to drift and scanning distortion, along with extending the field of view. The method allows flexible tuning of the spatial resolution and decouples the choice of field of view from the need for local atomic resolution. It consists of the acquisition of a series of independent small subimages containing an atomic resolution image of the local lattice. All subimages are then analysed individually for strain by fitting a nonlinear model to the lattice images. The obtained experimental strain maps are quantitatively benchmarked against the Bessel diffraction technique. We demonstrate that the proposed scanning strategy approaches the performance of the diffraction technique while having the advantage that it does not require specialized diffraction cameras.

**Introduction**

Moore's law describes a doubling of the amount of transistors per chip every year with later revisions to a doubling every two years [1]. This trend continued for quite a few decades until the discovery of reliability issues when reaching the nanoscale[2–4]. To tackle this problem, strain was introduced into the transistor channel to increase charge mobility, and thus transistor speed. Indeed, strain modifies the band-structure of silicon, germanium and their alloys: compressive strain increases hole mobility in p-MOS transistors and tensile strain increases electron mobility in n-MOS transistors[5,6]. Strain also increases the effective carrier velocity in the channel and improves the transport properties by compensating for the adverse effects of parasitic resistance and voltage reduction in short channel transistors [7]. Such properties give rise to large enhancement in the drive current [8] and significantly improves reliability when scaling down. Hence, monitoring strain in the production process with great precision, accuracy and spatial resolution is crucial for the semiconductor industry. Since there is no single consensual or universal technique to measure strain at the nanometer scale, one relies on potential inline techniques like Raman spectroscopy [9] in combination with offline techniques like X-ray strain microscopy [10] and transmission electron microscopy (TEM) to validate the results.

Although these techniques offer high strain precision (0.01%), they do not fulfil the demand for local strain measurement for the latest node of semiconductor devices due to their limited spatial resolution (100-500 nm)[11,12]. High resolution scanning transmission electron microscopy (HR-STEM) imaging is one of the TEM techniques with the capability to resolve individual atomic columns of a crystal and in combination with Geometric Phase Analysis offers spatial resolution for strain measurement up to a single unit cell [13,14]. Image patterns obtained with HR-STEM are relatively insensitive to thickness of the sample and can also work with slightly thicker samples (~200 nm) as opposed to conventional HRTEM techniques, which require thinner samples ~50 nm. The technique is also very fast in terms of data acquisition and analysis.

Unfortunately, HR-STEM images are plagued by scanning distortions and sample drift due to thermal or mechanical instabilities, resulting in artefacts in the strain measurement[15]. The scanning distortions arise from instabilities like spatial drift, electromagnetic interference or vibration and from the "flyback effect", when the electron probe scanning system needs time to move from one end of the line to the starting of the next line in a conventional raster scan scheme. This results in image distortions which are particularly problematic when measuring the subtle effect of strain on the lattice image. Several image correction techniques have been proposed in the literature which depend on multi-frame scanning and using rigid or non-rigid registration to average the multiple frames and estimate the deformation of the images [16,17]. The pixel dwell time used in each frame is rather short (≈ 0.5-1 µs) and the field of view is limited due to the required sampling of the crystal lattice and the limited amount of pixels in the image [18]. Exploiting a priori assumptions, Jones and Nellist were able to correct for scan noise and drift in STEM images [19]. Sang and LeBeau have shown that near perfect STEM images can be reconstructed using the RevSTEM technique[20], where a series of rotated images of the same area are used to estimate the drift rate and direction without prior information. These methods require multiple acquisitions of images resulting in a higher incoming electron dose (risk of beam damage), longer recording time and rely on the assumption that the drift is constant in time.

Alternative scanning strategies using spiral scans were also investigated by Sang. et.al. which eliminates the flyback effect common in conventional raster scanning. This offers very fast acquisitions but suffers from non-uniform sampling of the image. This can lead to blurring of images near the edges and a loss of contrast [21]. Ophus et.al. have proposed a linear and nonlinear drift and scanning artefact correction using two orthogonally scanned images [22]. This method is based on the near perfect information transfer in the fast scan direction but requires significant processing steps and sufficient low frequency information like edges or alignment markers for atomic resolution images acquired along low-index crystallographic zone axes. In this paper, we propose an alternative scanning strategy termed 'block scanning', where the probe scans the sample in a block by block fashion. Each block consists of a patch of atomic resolution image of the sample which can be used to evaluate the local strain in that region. This reduces the drift distortion which is prominent in a conventional raster scan as each block can be assumed to be near drift free given the much shorter recording time of an individual block. This method decouples the field of view from the local sampling requirements in each block and therefore allows a wider and freely adjustable field of view. We experimentally validate the proposed technique on a 16 nm Si-Ge Finfet device[23].

**Block scanning strategy**

High-angle annular dark-field imaging (HAADF) is a scanning transmission electron microscopy (STEM) technique which produces a dark field image formed by an electron probe of relatively large semi-convergence angle (~20 mrad). The image formed with this technique is proportional to the atomic number of the atoms, also called Z contrast imaging, and the sample thickness. In a raster scanning mode HAADF-STEM, the beam scans over the sample in a row by row fashion. This creates a fast scan direction for the rows and slow scan direction for the columns. Due to instabilities like spatial drift, electromagnetic interference or vibration, the lattice spacing information gathered is typically much more reliable in the fast scan direction as opposed to the slow scan direction. A modified scanning strategy is employed here, where the beam scans individual subimages consisting of a low number of pixels, each covering only a small fraction of the total exposure time (Figure 1b). This significantly reduces the delay in the slow scan direction and hence results in local lattice measurements that are significantly more robust against sample drift. Each individual subimage provides reliable lattice information in both the fast and slow scan direction. As long as the subimages are treated independently from each other, slow drift phenomena (> 100 ms) have no significant effect as will be demonstrated in the results section. Note that also the flyback issues are significantly improved as the total distance the probe has to move is now reduced to the distance between individual subimages. There however, is a large jump from the last subimage in the row to the left subimage in the next row which distorts a few pixels in that subimage. Hence, we do not consider the first column of subimages for the analysis. This problem can be mitigated by taking a smaller jump to the subimage in the next row by reorganising the order of the scanned blocks.

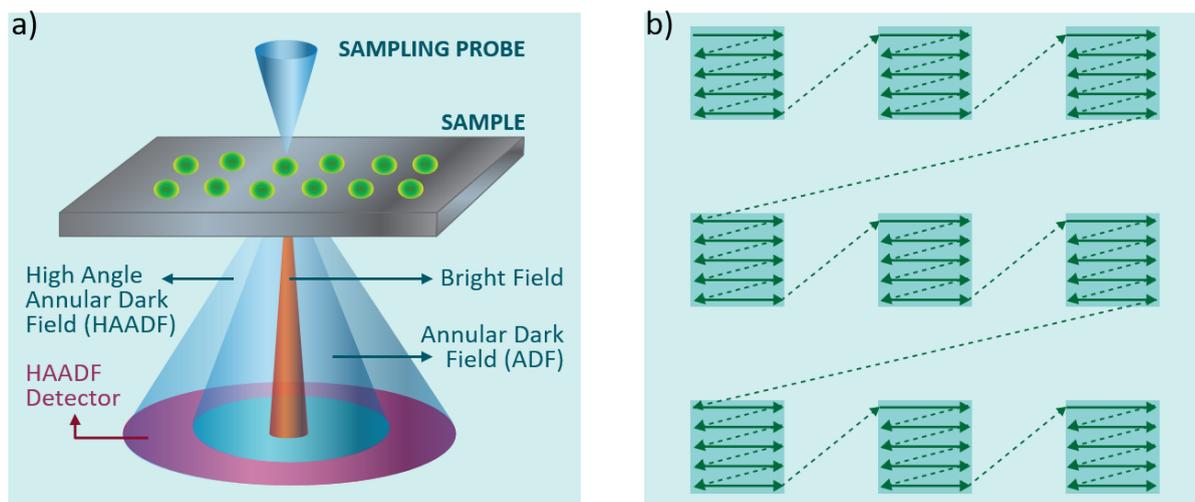

*Figure 1 a) In STEM imaging mode, the probe is positioned on the specimen at particular locations and the HAADF image intensity is recorded at every position b) Schematic representation of the block scanning strategy. The solid lines are the path taken during the scanning. The probe performs raster scan locally inside each block. The dashed lines indicate the spatial jumps taken by the probe while scanning.*

**Material**

The material of investigation is a 16 nm Si-Ge FinFET (Fin field effect transistor). The STI (shallow trench isolation) first approach is followed in device preparation. The device consists of a silicon substrate, which is etched to create shallow trenches and a thick $Si_{0.3}Ge_{0.7}$ layer is selectively epitaxially deposited (~110 nm) inside the trenches. The $Si_{0.3}Ge_{0.7}$ layer is completely relaxed due to the high aspect ratio [24] of the device and by the formation of dislocations and point defects at the silicon interface. This consequently acts as a strain relaxed buffer (SRB). A thin germanium channel (~30 nm) is grown epitaxially on top of the SRB and it conforms with the lattice parameter of the SRB. This channel is expected to be compressively strained in the lateral direction since the lattice parameter of $Si_{0.3}Ge_{0.7}$ is smaller than Ge. The fins are separated by silicon oxide amorphous material. The growth direction is [100] and the device is oriented in the {110} plane. The V-shape interface between the $Si_{0.3}Ge_{0.7}$ and the Si interface is used to trap the defects or dislocation from propagating into the Ge channel [25]. For the TEM analysis, two Focussed ion beam (FIB) samples were prepared along the cross section and the long section of the material of approximately 150 – 200 nm in thickness (Figure 2b and 2c) normal to <110>.

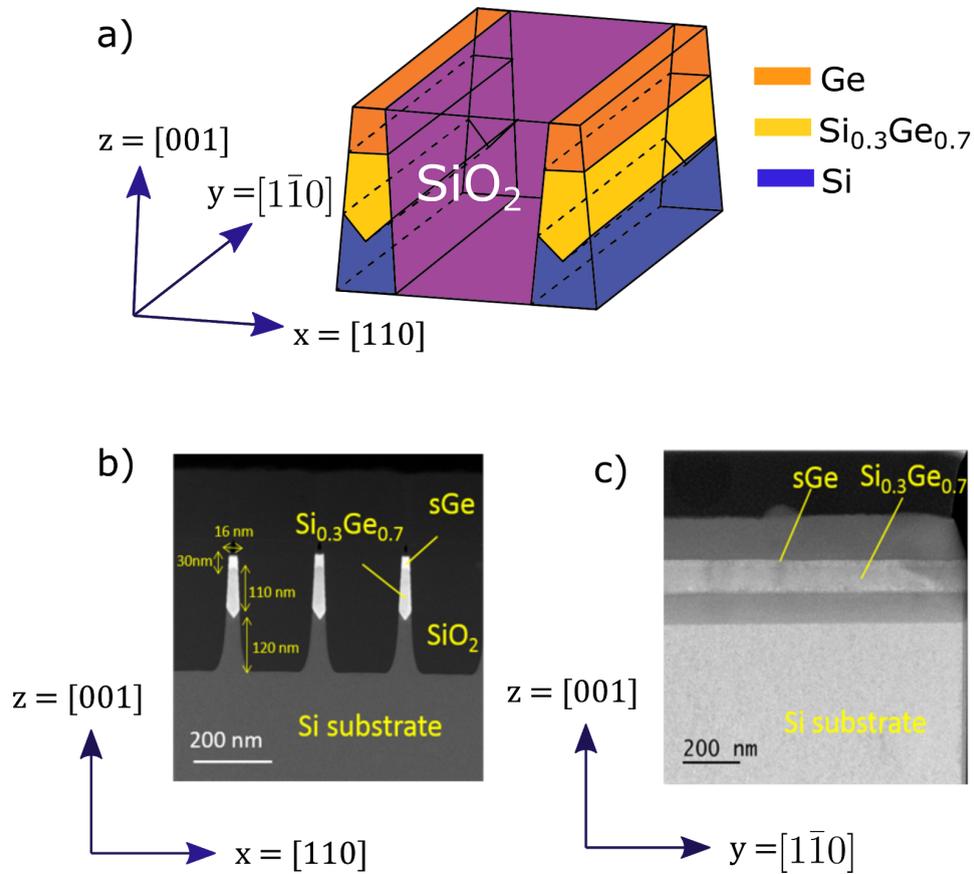

Figure 2 a) Schematic of the 16 nm FinFET b) cross section of the sample, extracting the xz plane c) Long section, extracting the yz plane.

**Experiment**

To implement the proposed idea in a transmission electron microscope, a custom scanning engine [collaboration M. Tence and M. Kociak [26,27]] is used to dynamically control the probe position. Full software control over the scanning grid is obtained on an aberration corrected FEI Titan microscope operating at 300 kV. The subimage sampling can be chosen by fine tuning the scan step size (distance between the probe visiting positions or the pixel size) and the number of pixels. The field of view is determined by the number of subimages and the distance between them, and can be extended up-to 250 - 300 nm as long as the Nyquist sampling criterion is satisfied inside each block. Figure 3 is an illustration of the block scanning technique on the long section TEM sample of the 16 nm FinFET. The sample was oriented in the [110] zone axis. The subimage sampling is tuned so that each block consists of 4 unit cells (1.5 x 2.2 $nm^2$, Figure 3b) spanning a total area of 192 x 282 $nm^2$. There is a spacing of one block (1.5 nm in y and 2.2 nm in z) between each block. Each block consists of 32 x 32 pixels, with a pixel size in y ≈ 0.047 nm and z ≈ 0.069 nm (rectangular pixel). This rectangular scanning grid is used to satisfy the Nyquist criteria in each block in agreement with the rectangular lattice. Alternatively, we have used a 64 x 64 pixel block, yielding a spatial resolution (block size) of 3 nm and a total field of view of 192 x 192 $nm^2$ with no spacing between the blocks, using square sampling, as the Nyquist criterion for this field of view was easier to satisfy. The total number of pixels used was 4096 x 4096 in both cases resulting in 64 x 64 subimages. We will compare the performance of both sampling choices further on. We use a dwell time of 20-50 μs to make sure all settling time issues with probe positioning are stabilised and a good SNR is available.

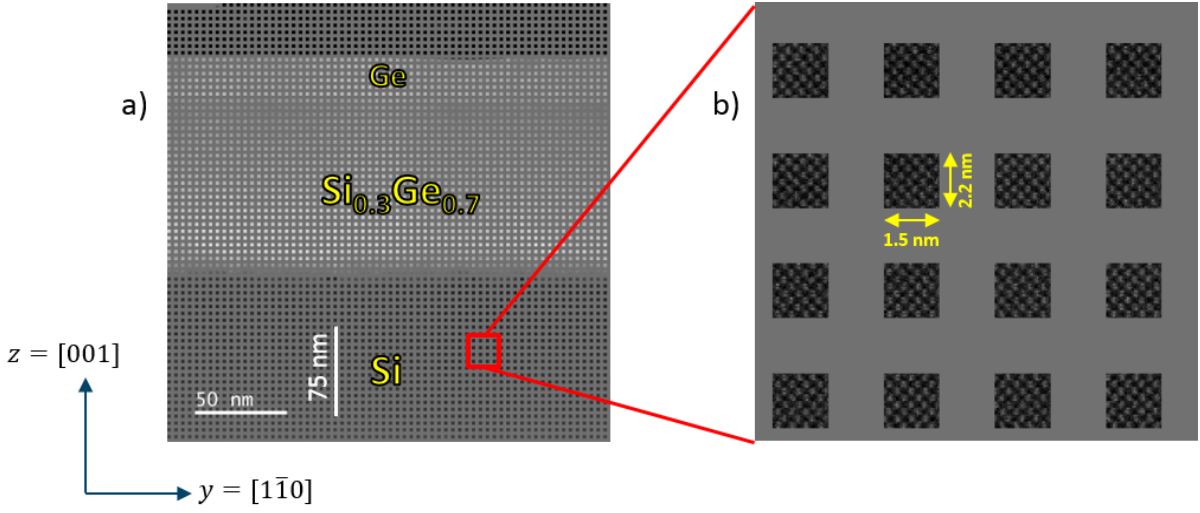

*Figure 3 a) Experimental block scanning STEM image of a 16 nm FinFET in the long section b) showing individual blocks in the Si region and each block consists of 32x32 pixels and covers approximately 4 unit cells.*

The goal now is straightforward to extract lattice parameters from each of the individual subimages. Since there is a limited number of pixels inside each block, conventional techniques like Geometric Phase Analysis (GPA) cannot be applied because GPA relies on information in the Fourier transform (FT) of the block. With 64 x 64 pixels, the FT lacks the frequency resolution to estimate strain accurately and a more sophisticated approach is necessary to obtain high frequency resolution in the Fourier domain with a limited number of pixels or samples. Hence, a parametric approach is taken where a model-based fitting or regression is used to fit the image data in real space inside each block.

**Non-linear 2D model fitting and strain calculation**

The Levenberg-Marquardt nonlinear least-square minimization [28] algorithm is applied for data fitting. The parameters of the 2D model extracted using this regression method are used to compute the value of strain inside each sub-image. Each block consists of a few atom columns (unit cells) and the atoms have a periodic lattice arrangement. This periodic lattice can be decomposed into a discrete sum of 2D sinusoids with only a few harmonics, analogous to a Fourier series decomposition of a periodic function. The initial model F(x,y) is constructed by computing the Fourier transform (FT) of each block. This gives an indication of the expected sinusoidal components representing each block (Figure 4b) but suffers from the fact that the small image patches are not strictly periodic (even though the underlying crystal is) because of the incommensurate cut at the image boundaries.

$$F(x,y) = A_0 + \sum_{n=0}^{N} \sum_{m=0}^{M} A_{m,n} \sin\left(2\pi(nf_x x + mf_y y) + \varnothing_{m,n}\right) \qquad (1)$$

Where, $A_0$ is the background intensity, $A_{m,n}$ is the amplitude of the 2D sinusoids, $f_x$ and $f_y$ are the spatial frequencies in the Cartesian coordinate system, $\varnothing_{m,n}$ is the phase of the sinusoids and M and N are the maximum number of harmonics in the model. Since the model is nonlinear in $f_x$, $f_y$ and

∅, the least square minimization algorithm requires a good initial guess to avoid getting stuck in a local minimum. The amplitude and phase of the sinusoidal components are estimated initially from the discrete 2D FT at the specific location of each diffractogram spot. The background value is estimated from the mean intensity value in the subimage. An illustration of the initial guess and a subimage after fitting using the Levenberg-Marquardt algorithm is shown in Figure 4c and 4d for a 64x64 pixel block.

The frequency vector $G = (f_x, f_y)$ of the 2D sinusoids is inversely proportional to the lattice spacing of the crystal and gives a direct indication of the lattice strain after comparison with the reference frequency vector $G_0$. The 2D distortion matrix D [29] is given by

$$D = \left(G^T G_0\right)^{-1} - I \qquad (2)$$

Where, $I$ is the identity matrix and $G^T$ is the transpose of $G$. The strain ε and rotation ω are calculated as

$$\varepsilon = \tfrac{1}{2}\left(D + D^T\right) \qquad (3)$$

$$\omega = \tfrac{1}{2}\left(D - D^T\right) \qquad (4)$$

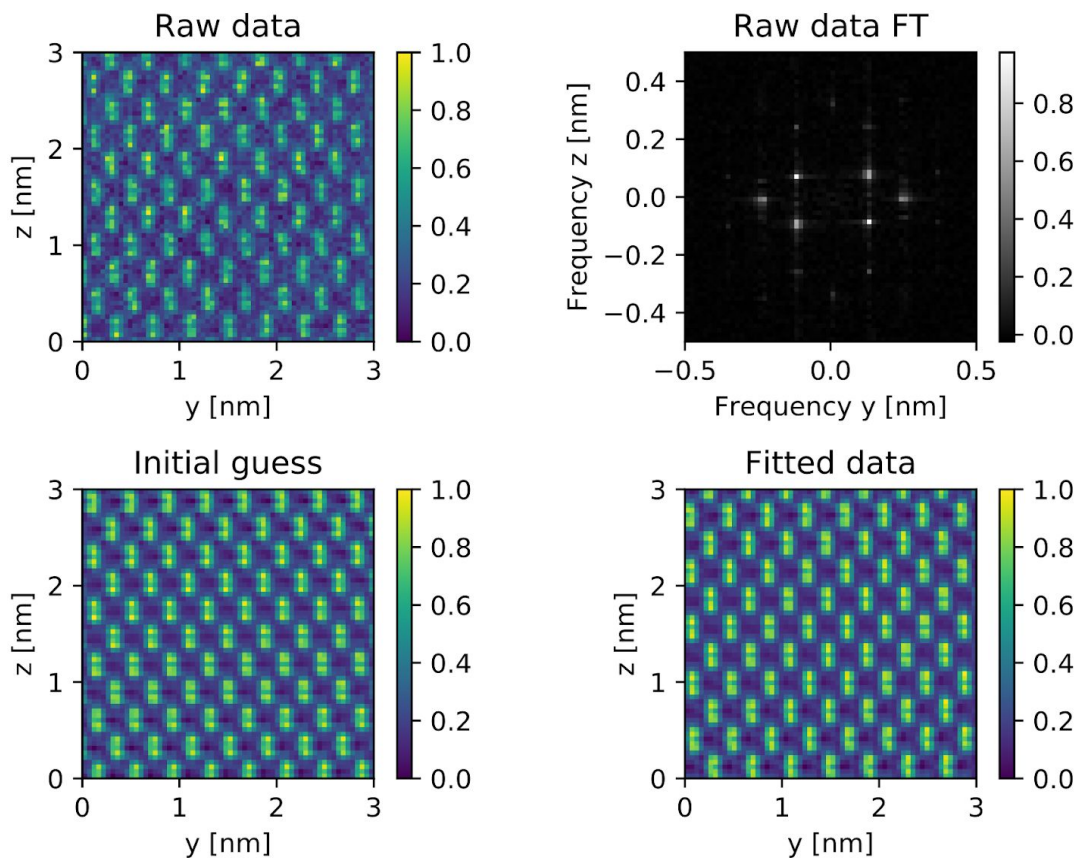

*Figure 4 a) Raw data of a block and its corresponding FT b). The amplitude of the frequency spectrum is shown with the low frequency component or the background removed to highlight the higher order frequency components, since the higher order components are masked with the high intensity background c) Initial guess computed from the FT of the raw data and*

*d) fitted block using Levenberg-Marquardt non-linear least squares minimization. The intensities of the data are normalised to have the same scale and better visualisation.*

The proposed block scanning technique was performed at 300 keV and with a beam current of 85 pA in the FEI Titan[3] aberration corrected microscope and is compared against the Bessel diffraction technique which was demonstrated to provide a precision and accuracy better than $2.5 \times 10^{-4}$ and $1.5 \times 10^{-3}$ respectively [30]. We also acquired HRSTEM images under the same imaging conditions as for the block scanning with dwell time 24 µs, field of view 192 x 192 nm$^2$ at 300 kV, on an unstrained Si sample to compare the strain precision in both the slow scan and the fast scan directions. Bessel diffraction was performed at 300 kV also in an aberration corrected FEI Titan[3] microscope. The diffraction pattern consists of overlapping rings (Figure 5) and is somewhat analogous to precession electron diffraction (PED) [31]. A hollow cone illumination is realised, where a simultaneous illumination of rays from different directions takes place as opposed to the sequential incidence in PED. This minimizes the hardware complexity of PED and the diffraction data is treated with dedicated python software, based on the autocorrelation of each diffraction pattern and subsequent extraction of the centre of the rings after background extraction and normalization [32,33]. The peak positions are calculated in reciprocal space and similar formulations as eqn. (2) to (4) are used for the strain calculation and mapping. The convergence angle was set to ~6 mrad to reduce the ring overlap area, which improves the precision of the measurement.

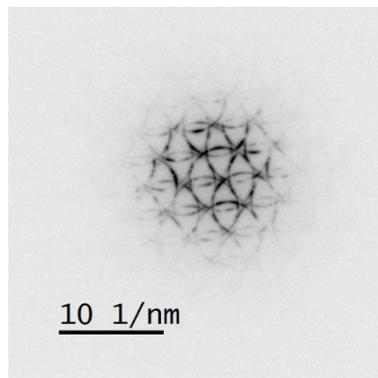

*Figure 5: Contrast reversed ring diffraction pattern obtained from the Bessel diffraction at 300 kV.*

**Results**

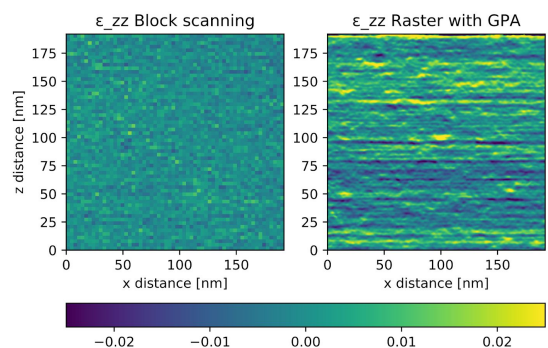

*Figure 6 Comparison of strain performance between block scanning and conventional raster scanning along the slow scan direction on an unstrained Si reference sample. The strain map in the raster scan is obtained using GPA*[29]*. The data*

Figure 6 shows a comparison of the proposed block scanning technique with the conventional raster scanning for a dwell time of 24 µs scanned over exactly the same area of a reference Si sample which is assumed strain free. For the raster scanned image, GPA analysis [29] is used to extract the strain values. Streaking artefacts in the horizontal direction are seen in the raster strain map due to misalignment between the adjacent rows originated from sample drift affecting the slow scan direction. In both cases the spatial resolution is 3 nm, determined by the selected GPA mask for raster scanning and by subimage size in the block scanning approach. The standard deviation on the strain observed in the slow scan direction for the block scanning is $3 \times 10^{-3}$ while it is $12 \times 10^{-3}$ for raster scanning. This demonstrates that up to fourfold improvement in the precision is possible using the proposed block scanning technique. In the fast scan direction, the precision values were close to each other with $1.2 \times 10^{-3}$ for the block scanning and $1.9 \times 10^{-3}$ for the raster scan with GPA.

Strain in classical mechanics is defined as a measure of deformation (contraction or elongation) of a material with respect to its original dimension or to a reference material. Experimental strain maps analysed with a 64 x 64 pixel block configuration on the 16 nm FinFET in the cross section and the long section configuration are shown in Figure 7 with respect to the underlying Si substrate as reference. The two perpendicular configurations of the lamella are used to estimate strain in all three perpendicular directions, assuming that the effect of relaxation due to specimen thickness can be neglected. Spatial maps and line profiles are compared with the Bessel diffraction results. The strain value in the Ge region (initially computed with Si as reference) is converted to bulk Ge as reference, because the actual strain with respect to the same Ge material is of importance. Table 1 shows the strain values in the Ge region calculated after conversion to the bulk Ge reference. Analysing the values reveals that the strain is completely relaxed in the x direction (across the fin), which can be understood due to the small width of the fin. There is a compressive strain along the fin in the Ge region which can be understood from the lattice mismatch between the $Si_{0.3}Ge_{0.7}$ SRB and the Ge channel. Due to the compression of the lattice in the y direction, there is a small tensile strain in the vertical direction in z, due to the Poisson effect [34].

Table 1 The strain values measured in the Ge region for the 16 nm finFET in the three perpendicular directions with respect to bulk Ge as the reference. The uncertainty values are the standard deviation in the measured area highlighted in red in the HAADF image in Figure 7 and represent the precision per block or per scanned point in the Bessel case.

| Technique | $\varepsilon_{xx}$ | $\varepsilon_{yy}$ | $\varepsilon_{zz}$ |
|---|---|---|---|
| Block scanning | $(-3 \pm 2) \times 10^{-3}$ | $(-15 \pm 2) \times 10^{-3}$ | $(6 \pm 3) \times 10^{-3}$ |
| Bessel | $(-2 \pm 1) \times 10^{-3}$ | $(-14 \pm 2) \times 10^{-3}$ | $(2 \pm 3) \times 10^{-3}$ |

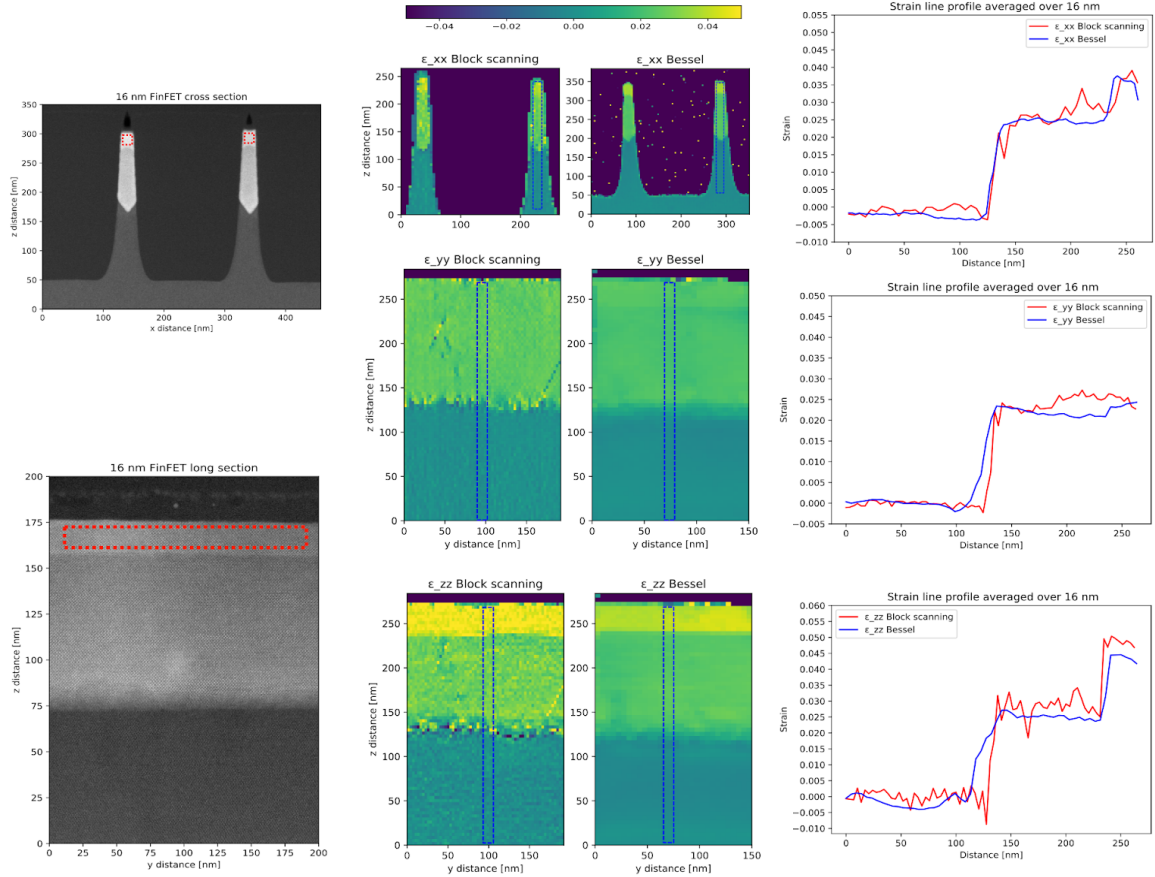

*Figure 7 Strain maps on the 16 nm FinFET in the cross section and the long section and a comparison with the Bessel diffraction technique with respect to the Si substrate. The line profiles are averaged over 16 nm as indicated in the dashed blue box in the strain maps. The HAADF image on the left shows the cross section and the long section of the 16 nm FinFET and the Ge area is highlighted in red.*

The strain precision per block was measured as the standard deviation over the Si reference and for the block scanning the precision is $1 \times 10^{-3}$ in the fast scan direction and $3 \times 10^{-3}$ in the slow scan direction. The precision is slightly lower in the slow scan direction due to drift and distortions that can still take place during the line scan inside a block, but the effects are reduced drastically in comparison to the conventional raster scan (with a four fold improvement as demonstrated in Figure 6). The precision decreases by three-fold with the use of 32 x 32 pixel blocks due to the smaller number of pixels used to estimate the parameters of the model. Velazco et. al have demonstrated different scanning strategies like 'Snake' and 'Hilbert' which results in decreasing the slow scan distortion by eliminating the preferential slow scan direction altogether [35]. However, some effects of sample drift distortion are still present at higher pixel acquisition times, if the strain analysis is done on the whole field of view instead of independent block by block fashion.

The strain precision for the Bessel diffraction was estimated at $4 \times 10^{-4}$ per scanned point from the acquired data. The accuracy of both the Bessel and block scanning techniques are equal at $2 \times 10^{-3}$, which was derived from a bulk Ge sample (known lattice parameter $a_{Ge} = 0.56\ nm$) with Si as a reference assuming no relaxation due to thickness effects in these reference areas. To test the best possible results that could be achieved with the block scanning technique, HAADF multislice simulation was performed to generate hypothetical noise free HAADF images that are devoid of distortions of the scan system or sample drift (Figure 8a). The aberration values were chosen close to the experimental conditions with defocus = 2 nm, two fold astigmatism A1 = 966.8 pm and spherical aberration Cs = -274.8 nm. The parameters were estimated using the probe corrector software at the start of the experiment. Images of a layered strained structure of Si are simulated by artificially varying the lattice parameter with strain values varying from -0.05 up to 0.05 on top of an unstrained Si substrate. The size of individual blocks was set to 64 x 64 pixels (Figure 8). The simulated data was then treated with the same optimisation algorithm as used for the experimental images.

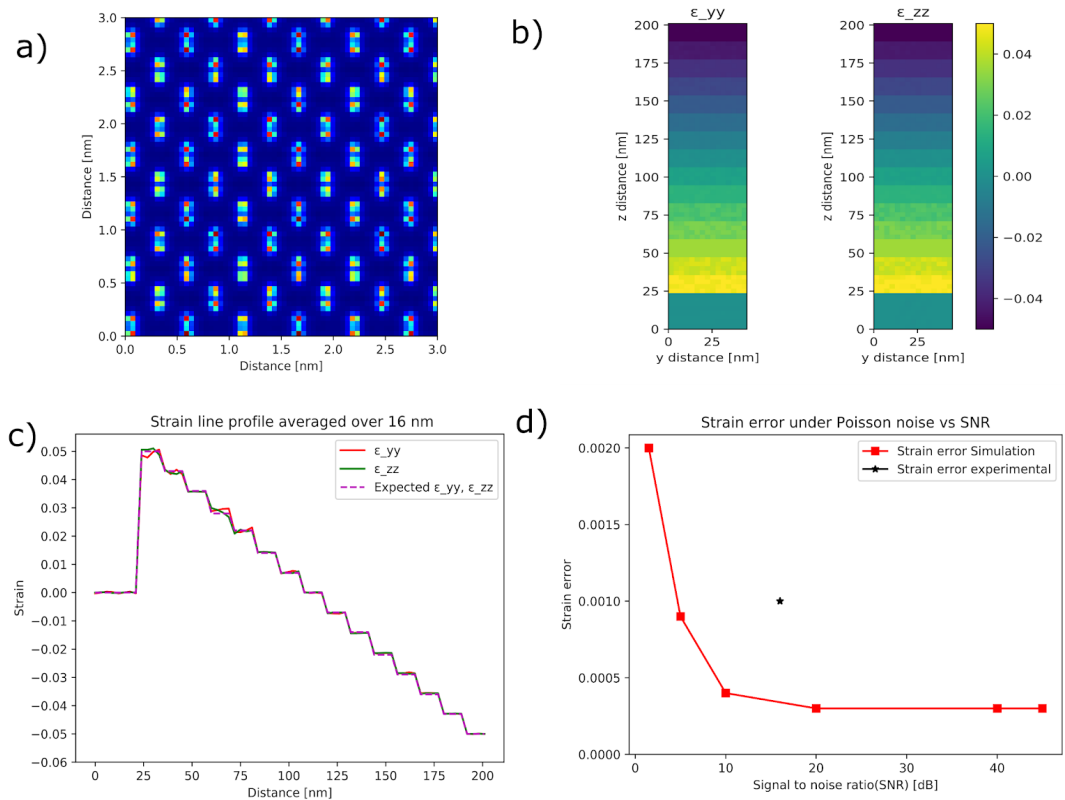

*Figure 8 a) Simulated block of 64x64 size and 3x3 nm² area b) Strain map showing the simulated layered structures with strain profiles from -0.05 up to 0.05 on top of an unstrained structure c) Line profile obtained vertically on the strain map averaged horizontally across 16 nm d) Strain error, which, is a combination of precision and bias, calculated as a function of Signal to noise ratio (SNR) under the addition of Poisson noise.*

The best possible precision and accuracy calculated on the simulated subimage data set of 64 x 64 pixels (spatial resolution/subimage size of 3 nm) are $2 \times 10^{-4}$ and $4 \times 10^{-4}$ respectively. These values are highly attractive for strain mapping in nanostructures.

The small amount of bias leading to a finite accuracy is due to the simplified model that represents the HAADF image with a sum of a few sinusoids with their harmonics and might include numerical uncertainties. It is observed that there is a slight bias in the strain value depending on the position of the crystal lattice in the subimage. This bias decreases significantly when increasing the number of pixels in a subimage and varies also with the number of unit cells inside the block or indirectly to the periodicity of the sinusoids. This is confirmed by simulating two data sets: 1) 32 x 32 pixels block with 1.5 nm spatial resolution (block size) and a 2) 64 x 64 pixels block with 3 nm spatial resolution. The accuracy improves significantly from $2 \times 10^{-3}$ in the former case to $4 \times 10^{-4}$ in the latter. In a realistic scenario, increasing the number of pixels in a subimage would also lead to more drift inside the block and this would experimentally lead to more distortions in the slow scan direction. Hence, there is a trade-off between the number of pixels inside a block (and the spatial resolution) and the attainable precision. In addition, Poisson noise was added to the simulated subimage to mimic the counting noise of the HAADF detector (Figure 8d). It can be noticed that the precision scales, as expected, as a square root function of the number of counts, with a plateau setting in from an SNR of about 20 dB. This indicates that the systematic errors are introducing a bias from this point on. The bias can be due to sampling, model inaccuracies and numerical implementation, and should not influence the result more than 4 x $10^{-4}$.

For the experimental data in Figure 7, we choose a beam current of 85 pA at 300 kV during image acquisition. A convergent beam electron diffraction (CBED) pattern is simulated and the expected number of electrons $N$ falling on the HAADF detector (with an inner collection angle of 44 mrad and outer angle of 190 mrad) is estimated for the given current and the acquisition time of 24 μs. The SNR is calculated as $\sqrt{N}$ following the Poisson statistics, which yields an SNR of 16 dB. At this level of SNR, we expect the precision to be dominated by the remaining bias issues and are no longer SNR limited. Moreover, in the practical realisation of the experiment, more sources of bias exist, such as sample drift and probe control inaccuracies that will further increase the effect of bias on the obtained strain values.

To finally evaluate the model, a normalised root mean square error (NRMSE) between the fitted model and the experimental data of a 16 nm finFET(Figure 7) long section (with 1 being no fitting and 0 being the exact fit) is shown in Figure 9.

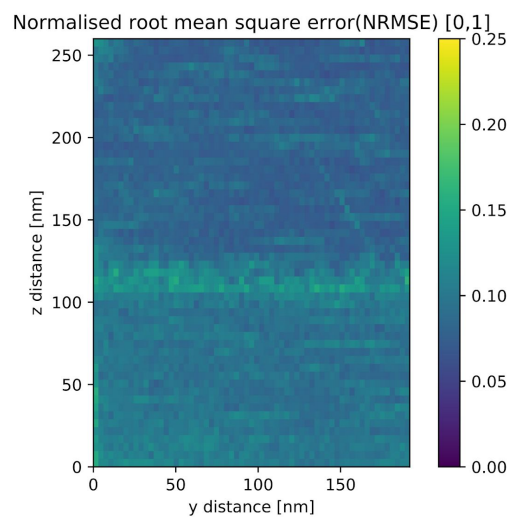

*Figure 9 Normalised Root Mean square residual(NRMSE) of the 16 nm FinFET long section in the range [0,1], after fitting each block with the model. Some features exist near the interface, indicating that here, our assumption of uniform strain in the sample thickness direction is not valid. The mean and standard deviation of the NMRSE are 0.09 and 0.04 indicating a good fit with the model. There is no particular trend seen in the plot (apart from the interface) indicating that the residuals are indeed random and the model is a good representation of the data.*

We observe a higher residual near the Si - $Si_{0.3}Ge_{0.7}$ interface due to reduced contrast in the subimages, related to the projection of regions where strain is not uniform across the thickness of the sample. Such non-uniformity breaks the alignment of the atoms along the direction, leading to lower 'high resolution' contrast. Apart from this region, the residual is featureless indicating an adequate model for the data. With the current implementation, the fitting time takes about 1 second per 64x64 block, resulting in a computation time per image of about 66 minutes. This time could be drastically reduced by further optimizing the code.

**Conclusion**

We document the implementation of a new block scanning method for local strain measurement in TEM. This technique efficiently minimizes drift and slow scan distortions that are typically hindering the use of HRSTEM imaging for strain measurement. Moreover, the method provides a larger and flexible field of view in comparison to conventional HRSTEM imaging. An analytical harmonic model is constructed for each individual subimage starting from initial parameters obtained from its Fourier transform. Using model based non-linear fitting, the parameters of this model are estimated and reveal the local strain in each block independently. This technique is demonstrated on a 16 nm FinFET device, and the measured strain is compared with the Bessel diffraction setup. Although the precision obtained with the block scanning technique is comparable to HRSTEM combined with GPA and slightly below what is attainable with Bessel diffraction, the technique offers a much wider field of view in comparison to conventional HRSTEM and requires no specialised diffraction cameras.

Acknowledgements.

A.B. D.J. and J.V. acknowledge funding through FWO project G093417N ('Compressed sensing enabling low dose imaging in transmission electron microscopy') from the Flanders Research Fund. J.V acknowledges funding from the European Union's Horizon 2020 research and innovation programme under grant agreement No 823717 – ESTEEM3. The Qu-Ant-EM microscope and the direct electron detector used in the diffraction experiments was partly funded by the Hercules fund from the Flemish Government. This project has received funding from the GOA project "Solarpaint" of the University of Antwerp. GG acknowledges support from a postdoctoral fellowship grant from the Fonds Wetenschappelijk Onderzoek - Vlaanderen (FWO). Special thanks to Dr. Thomas Nuytten, Prof. Dr. Wilfried Vandervorst, Dr. Paola Favia, Dr. Olivier Richard from IMEC, Leuven and Prof. Dr. Sara Bals from EMAT, Antwerp for their continuous support and collaboration with the project and to the IMEC processing group for the device fabrication.